\documentclass[
reprint,
superscriptaddress,
nofootinbib, 
 amsmath,amssymb,
 aps
]{revtex4-1}

\usepackage{graphicx}
\usepackage{bm}
\usepackage{hyperref}
\usepackage{color}
\usepackage{graphicx}
\usepackage{multirow}
\usepackage{ulem}

\usepackage{academicons}
\usepackage{xcolor}
\newcommand{\orcid}[1]{\href{https://orcid.org/#1}{\includegraphics[scale=0.5]{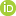}}}

\newcommand{\beq}{\begin{equation}}
\newcommand{\eeq}{\end{equation}}
\def\bea{\begin{eqnarray}}
\def\eea{\end{eqnarray}}
\def\beal{\begin{align}}
\def\eal{\end{align}}
\newcommand{\bei}{\begin{itemize}}
\newcommand{\eei}{\end{itemize}}
\newcommand{\bmat}{\begin{matrix}}
\newcommand{\emat}{\end{matrix}}

\newcommand{\Fig}[1]{Fig.~\ref{#1}}
\newcommand{\Eq}[1]{Eq.~(\ref{#1})}

\newcommand{\Sec}[1]{Sec.~\ref{#1}}

\def\={\,=\,}
\def\+{\,+\,}
\def\-{\,-\,}

\def\Msun{M_\odot}

\begin{document}

\title{
Importance of eccentricities in parameter estimation of compact binary inspirals with decihertz gravitational-wave detectors
}

\author{Han Gil Choi~\orcid{0000-0002-1390-9024}}
\email{hgchoi1w@gmail.com}
\affiliation{Cosmology, Gravity and Astroparticle Physics Group, Center for Theoretical Physics of the Universe,
Institute for Basic Science (IBS), Daejeon, 34126, Korea}

\author{Tao Yang~\orcid{0000-0002-2161-0495}}
 \email{yangtao@whu.edu.cn}
 \affiliation{School of Physics and Technology, Wuhan University, Wuhan 430072, China}
\affiliation{Center for the Gravitational-Wave Universe, Astronomy Program Department of Physics and Astronomy, Seoul National University, 1 Gwanak-ro, Gwanak-gu, Seoul 08826, Korea}
\author{Hyung Mok Lee~\orcid{0000-0003-4412-7161}}
 \email{hmlee@snu.ac.kr}
 \affiliation{Center for the Gravitational-Wave Universe, Astronomy Program Department of Physics and Astronomy, Seoul National University, 1 Gwanak-ro, Gwanak-gu, Seoul 08826, Korea}

\date{\today}

\begin{abstract}
During its inspiral stage, a binary black hole (BBH) produces characteristic gravitational wave (GW) signals. The waveform of the GW signals can be described by the physical parameters of BBH, such as the masses of the black holes and the orbital eccentricity. Precise and accurate estimation of these parameters is crucial for GW astrophysics. In the aspect of precision, decihertz GW detectors are promising proposals, as they are anticipated to allow us to obtain highly precise parameter estimations for stellar-mass BBHs. However, the high-precision parameter estimation requires accurate GW waveform modeling. Otherwise, systematic errors can arise in estimated parameters. We emphasize the importance of considering the orbital eccentricity in constructing an accurate GW waveform model. B-DECIGO and MAGIS are used as benchmarks for decihertz GW detectors. We examine the significance of systematic error for a population of stellar-mass BBH inspirals. We found that the quasicircular GW waveform model exhibits significant systematic errors for BBH with a very small eccentricity $\sim 10^{-4}$ at GW frequency $0.1\, \text{Hz}$. The modeling accuracy can be substantially enhanced by incorporating the leading-order correction to GW phase evolution associated with eccentricity smaller than 0.01. The higher-order post-Newtonian corrections induced by eccentricity should be important only for eccentricity larger than 0.01.
\end{abstract}

\maketitle


\section{Introduction} \label{sec:intro}

Predictions on the actual eccentricity distribution of merging compact binaries depend on the formation scenario. There are two popular formation scenarios: isolated formation~\cite{Bethe:1998bn,Kowalska:2010qg,Dominik:2012kk,Bavera:2020uch} and dynamical formation~\cite{Bae:2013fna,Rodriguez:2016kxx,Park:2017zgj,Hong:2015aba,Samsing:2013kua,Samsing:2017xmd,Rodriguez:2018pss,Zevin:2021rtf,OLeary:2008myb,Antonini:2012ad,Takatsy:2018euo,Tagawa:2020jnc,Gondan:2020svr,Tagawa:2020qll}. In the isolated formation, the binary orbital evolution is primarily driven by interactions with companion stars, and majority of binary mergers occur with very small eccentricities. In this scenario, BBH mergers with $e> 0.01$ at GW frequency $10\, \text{Hz}$ are expected to account for much less than $1\, \%$ of the total binary merger population~\cite{Kowalska:2010qg}. The dynamical formation is relevant to binary formation in globular clusters~\cite{Bae:2013fna,Rodriguez:2016kxx,Park:2017zgj,Hong:2015aba,Samsing:2013kua,Samsing:2017xmd,Rodriguez:2018pss,Zevin:2021rtf} and nuclear star clusters~\cite{OLeary:2008myb,Antonini:2012ad,Takatsy:2018euo,Tagawa:2020jnc,Gondan:2020svr,Tagawa:2020qll}. In globular clusters, binaries with high eccentricities can be produced by two-body encounters~\cite{Hong:2015aba} or binary-single encounters~\cite{Samsing:2013kua,Samsing:2017xmd}. Consequently, studies based on \textit{N}-body simulations~\cite{Rodriguez:2018pss,Zevin:2021rtf} estimated that binary mergers with $e>0.01$ at $10\, \text{Hz}$ account for $\mathcal{O}(1)$ percents of the total population. In nuclear star clusters, much higher eccentricities are expected due to gravitational capture~\cite{OLeary:2008myb} and the influence of supermassive black hole~\cite{Antonini:2012ad}.

It is expected that the majority of mergers will have small eccentricity at high GW frequencies. One of the reasons is that the eccentric orbit of merging binary tends to be circularized due to GW emission~\cite{Peters:1963ux,Peters:1964zz}. For example, if the eccentricity was $0.1$ at $f=0.1~\text{Hz}$, it decreases to $e \sim 0.001$ as the GW frequency increases to $f=10~\text{Hz}$, which can be estimated from $e \propto f^{-19/18}$~\cite{Peters:1964zz}. This eccentricity decay takes less than a year in the case of stellar-mass BBHs. As a result, GW observation in a lower GW frequency band is more relevant to eccentric mergers. 

It has been demonstrated that observing eccentric binary mergers in ground-based detectors such as LIGO and Virgo can provide evidence for the dynamical formation of binary systems~\cite{Bae:2013fna,Rodriguez:2016kxx,Park:2017zgj,Hong:2015aba,Samsing:2013kua,Samsing:2017xmd,Rodriguez:2018pss,Zevin:2021rtf,OLeary:2008myb,Antonini:2012ad,Takatsy:2018euo,Tagawa:2020jnc,Gondan:2020svr,Tagawa:2020qll}. 
Measuring the orbital eccentricity with the space-based GW detector LISA can help discriminate between different formation channels of stellar-mass binary black holes (BBH)~\cite{Nishizawa:2016jji,Nishizawa:2016eza,Breivik:2016ddj}. In addition, the precision of the sky localization of eccentric binary inspirals tends to be better than those of quasicircular cases in the ground-based detector networks~\cite{Sun:2015bva,Ma:2017bux,Pan:2019anf}. Recently, Refs.~\cite{Yang:2022tig,Yang:2022iwn,Yang:2022fgp} showed that the improvement of sky localization of eccentric binary inspirals is significantly more pronounced in the space-based detector operating in the decihertz band. Another promising application of eccentric binary inspirals is the early detection and localization of GWs by utilizing the high harmonic modes induced by eccentricity~\cite{Yang:2023zxk}.

The space-based interferometer DECIGO~\cite{Kawamura:2018esd} and the space-based atom interferometer MAGIS~\cite{Graham:2016plp,Graham:2017pmn} have been proposed to cover decihertz GW frequencies. We refer to detectors that are sensitive in the range of deci-Hz to several Hz range as midband detectors. The midband detectors have great potential for GW science with stellar-mass compact binaries. Binaries composed of stellar mass black holes can emit GW signals for weeks and months in the decihertz band. Continuous observation of GW signal from compact binaries for long duration enables us to achieve not only high signal-to-noise (SNR) ratio of the GW signal but also exquisite measurement of the GW source parameters~\cite{Nair:2015bga,Nair:2018bxj,Isoyama:2018rjb,Sedda:2019uro,Graham:2017lmg,Yang:2021xox}. Therefore, successful operation of the midband detectors will provide valuable hints to the formation of compact objects~\cite{Nakamura:2016hna} as well as the test of general relativity~\cite{Yagi:2009zz,Choi:2018axi}.

The high-precision measurement of parameters with midband detectors requires more accurate GW waveform modeling. Otherwise, systematic errors can significantly contaminate the parameter estimations of GW signals. Specifically, we test the accuracy of estimated parameters assuming the quasicircular waveform model for the eccentric merger of BBHs. This issue has been examined in the context of ground-based detectors~\cite{Martel:1999tm,Favata:2013rwa,Favata:2021vhw,Cho:2022cdy}. In Ref. \cite{Favata:2013rwa}, the author showed that the quasicircular waveform model becomes inaccurate for eccentric mergers with $e\sim 10^{-3}$ at $f=10~\text{Hz}$ in the case of neutron star binary merger. It has been also noted that extending the low-frequency limit results in observing more GW cycles, consequently amplifying the systematic errors~\cite{Favata:2021vhw}. Therefore, we expect that the issue of systematic errors will be more pronounced in the midband detectors. 

Waveform model constructed using the Post-Newtonain (PN) approximation always has the risk of systematic error coming from neglected high-order PN effects~\cite{Tanay:2016zog,Tanay:2019knc,Cho:2021oai}. Therefore, it is essential to ensure that the waveform model being used includes high enough PN order terms to minimize systematic errors. In our study, we focus on the PN corrections associated with eccentricity in the context of observation with the midband detectors.

The remainder of this paper is organized as follows. Section \ref{sec:setup} introduces the eccentric waveform models and the GW detectors used in our work.  Section \ref{sec:PE} provides brief reviews of the GW parameter estimation. Computation methods for probability distribution are also reviewed in this section. In Sec. \ref{sec:sys}, the methods for systematic error estimation are introduced. Section \ref{sec:sysresult} focuses on the systematic errors due to neglecting eccentricity or inaccurate eccentric waveforms. In Sec. \ref{sec:implication}, we discuss the observational implications of the systematic errors. Lastly, Sec. \ref{sec:conclusion} is the summary and conclusion of our work.

\section{Basic setup}\label{sec:setup}
\subsection{Waveform model}\label{sec:waveform}

In GW polarization basis, the GW with the polarization amplitudes $h_{+,\times}$ passing through the detector with antenna pattern functions $F_{+,\times}$ will induce the strain signal
\begin{equation}
h(t) = F_+ h_+(t) + F_\times h_\times (t)\, .
\end{equation}
We use the Fourier transform of the signal 
\begin{equation}
    \tilde{h}(f) \equiv \int_{-\infty}^{\infty} df h(t) e^{2\pi i f t} dt
\end{equation}
for the parameter estimation. The integral can be calculated via the stationary phase approximation (SPA). 
In terms of SPA amplitude $\mathcal{A}$ and phase $\Psi$, the Fourier transform of the waveform can be written as
\begin{subequations}\label{eq:hamp}
\begin{align}
        &\tilde{h}= \mathcal{A}e^{i\Psi}\, , \text{where}\\
        &\mathcal{A}=- M \sqrt{\frac{5\pi}{96}}\left(2\frac{M}{\overline{D}}\right)\sqrt{\eta}(\pi M f)^{-7/6}\, .\label{eq:amp}
\end{align}
\end{subequations}
Here, $M=m_1+m_2$ is the binary total mass in the detector frame, $\eta= m_1 m_2/M^2$ is the symmetric mass ratio, and
\begin{equation}
\overline{D} \equiv 2 d_L [(1+\cos \iota ^2)^2F_+^2+ 4\cos \iota ^2 F_\times^2]^{-1/2}
\end{equation}
is the effective distance that absorbs the dependence on the luminosity distance $d_L$ to source, the inclination angle of the binary $\iota$.
$F_{+,\times}$ depends on the detector orientation, the GW source direction, and the GW polarization. In GW observations with space-based detectors, the detector's orbital motion causes significant time variation in detector orientation, making $F_{+,\times}$ time-dependent quantities. However, for computational efficiency, we fix the values of $F_{+, \times}$. This approximation is justified because the time-dependent $F_{+,\times}$ effectively result in a slowly changing GW amplitude over a much longer timescale than the GW period ($\sim f^{-1}$), and thus, it does not affect the GW phase measurement, which is our primary concern in this work.

The SPA phase is composed of several distinct contributions~\cite{Buonanno:2009zt,Moore:2016qxz,Kim:2019ecc}
\begin{equation}
\begin{split}
    \Psi(f) = &\phi_c + 2\pi f t_c \\
    &+\frac{3}{128 \eta v^5}\big(1+\Delta\Psi^\text{circ}_\text{3.5PN}
    +\Delta\Psi^\text{ecc}_\text{3PN}\big)\, , \label{eq:phase}
    \end{split}
\end{equation}
where $\phi_c$ and $t_c$ are the coalescence phase and time, and $v=(\pi M f)^{1/3}$ is the post-Newtonian (PN) orbital velocity parameter. In this expression, we factor out the leading order inspiral phase $3/(128\eta v^5)$ and divide the inspiral phase corrections into circular term and eccentric term. The circular term $\Delta\Psi^\text{circ}_\text{3.5PN}$ is the PN correction of the quasicircular binary inspiral, and therefore it is nonvanishing even for zero-eccentricity orbits. It includes the PN corrections up to 3.5 PN order. The explicit expression of the $\Delta\Psi^\text{circ}_\text{3.5PN}$ can be read off ~\cite{Buonanno:2009zt}. The eccentric term accurate up to 3 PN order can be written as \cite{Moore:2016qxz}
\begin{align}\label{eq:eccph}
    \Delta\Psi^\text{ecc}_\text{3PN} = -\frac{2355}{1462} e_0^2 \bigg(\frac{v_0}{v}\bigg)^{19/3}\left[1+\sum_{n=2}^{6}c_n v^n\right]\, ,
    \end{align} 
where $v_0=(\pi M f_0)^{1/3}$ is the PN orbital velocity parameter at a reference GW frequency $f_0$, and the $e_0$ is the orbital eccentricity at GW frequency $f_0$. The reference GW frequency $f_0$ can be chosen arbitrarily, so we set $f_0=0.1\, \text{Hz}$. The coefficients $c_n$ are functions of binary black hole masses and GW frequency. The details of $c_n$ can be found in Ref.~\cite{Moore:2016qxz}.  

Our waveform model is a minimal setup designed to study eccentric GW waveforms. This approach allows us to reveal the effects of eccentricity more clearly. However, it is important to note that eccentric GW waveforms can be much more complicated in reality. For instance, an eccentric GW waveform consists of multiple harmonics of orbital frequency $f_\text{orb}$. While second harmonic $f=2f_\text{orb}$ predominates in the small eccentricity regime, the subdominant modes can potentially impact parameter estimation~\cite{Yang:2022tig,Yang:2022iwn,Yang:2022fgp,Moore:2019vjj}. In addition, the phase corrections from black hole spins, which we have not incorporated, introduce additional complications into parameter estimation~\cite{Isoyama:2018rjb}. This aspect remains for future investigation.

GW waveform defined by Eqs. (\ref{eq:hamp}), (\ref{eq:phase}) and (\ref{eq:eccph})  is the so-called TaylorF2Ecc~\cite{Kim:2019ecc}. 
In the following sections, we simulate GW signals from BBH sources with this waveform model. To study waveform accuracy, we define reduced waveform models EccMPN by collecting PN corrections in \Eq{eq:eccph} up to $v^{2M}$ terms. Since the $v^1$ term is absent, there is no Ecc0.5PN. We also consider the waveform model with $\Delta\Psi^\text{ecc}_\text{3PN}=0$ which is equivalent to TaylorF2~\cite{Buonanno:2009zt}; we refer to this model as QC for brevity. 

\subsection{GW detectors and the frequency range} \label{sec:detector}
\begin{figure}
\includegraphics[width=1\linewidth]{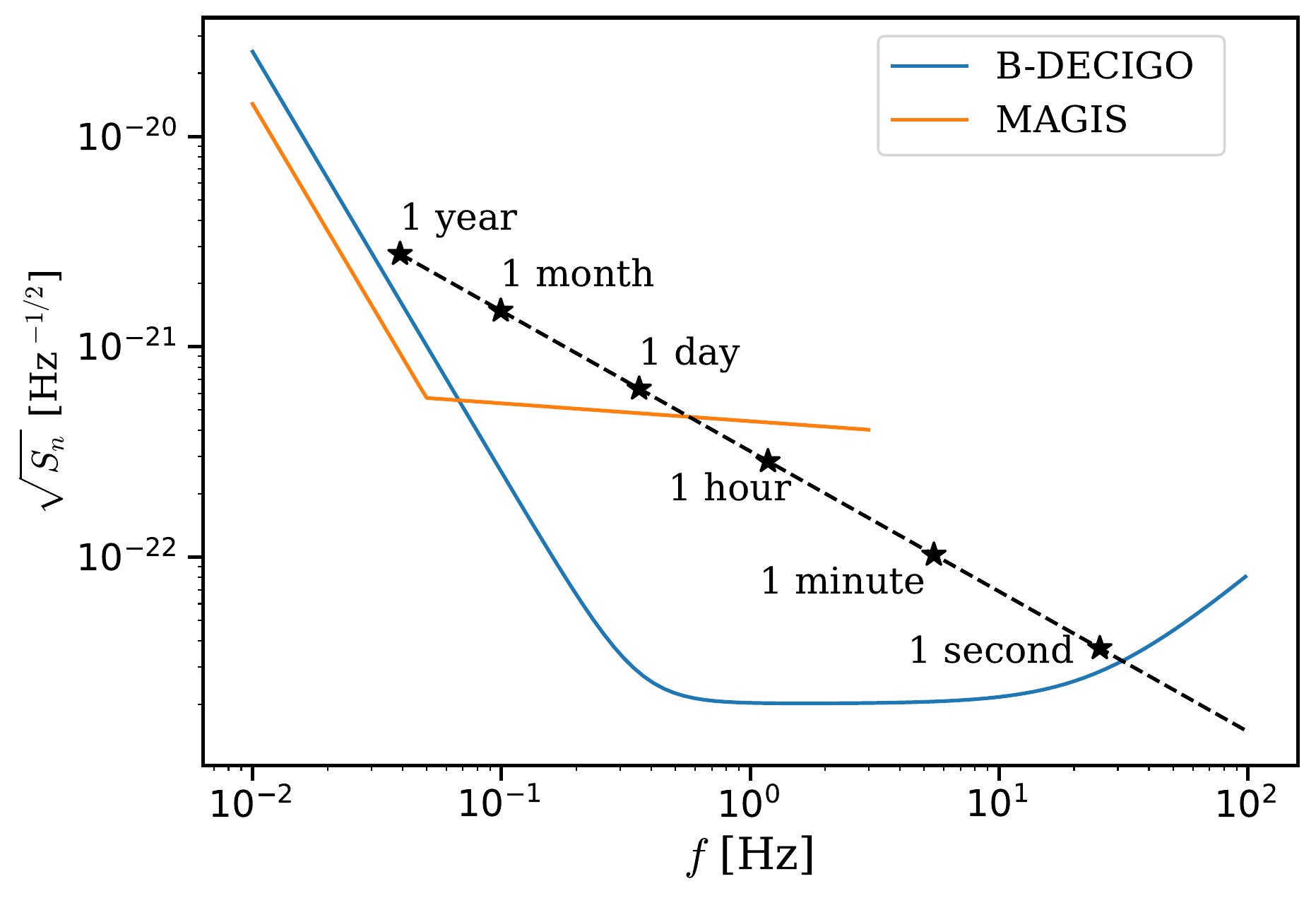}
\caption{ \label{fig:sensitivity} 
Sensitivity curves of B-DECIGO (blue) and MAGIS (orange). The square root of the noise PSD $\sqrt{S_n(f)}$ is used to present them. In principle, the decihertz band detectors have the capability to observe the one-year-long inspiral of stellar mass BBHs. The GW spectrum of BBH inspiral (dashed) with $M = 40 M_\odot $, $\eta = 0.24$, and $\overline{D}=5194~\text{Mpc}$ is presented as an example. The star markers with labels show the inspiral lifetime at the frequencies.
}
\end{figure}

We examine the near-future GW observations made by B-DECIGO \cite{Kawamura:2018esd} and MAGIS \cite{Graham:2016plp,Graham:2017pmn}.
 Both are space-based GW observatories and are sensitive to decihertz GWs. The sensitivity curves of the detectors are shown in \Fig{fig:sensitivity}. These detectors are anticipated to have exquisite precision in measuring the inspiral motion of BBH with stellar mass, as minute variations in GW phase evolution accumulate over the year-long inspiral lifetime in the decihertz band, allowing for measurability~\cite{Isoyama:2018rjb, Choi:2018axi}. Their precision may require more accurate source modeling than that needed in LIGO, which is one of the motivations of our study.

The B-DECIGO noise power spectral density (PSD) is obtained by the analytic fitting function provided in Eq. (20) of Ref. \cite{Isoyama:2018rjb}. In the case of MAGIS, we used
\begin{equation}
    S_n(f) = \begin{cases} S_* \left(\frac{f}{f_{*}}\right)^{-4} & f< f_* \\ S_* \left(\frac{f}{f_{*}}\right)^{-0.17}& f\geq f_* \end{cases}~,
\end{equation}
where $S_* = 3.26\times 10^{-43}~\text{Hz}^{-1}$ and $f_* = 0.05~\text{Hz}$, which approximately fits the resonant mode sensitivity curve in Ref. \cite{Graham:2016plp}. In our analysis, the frequency bands of B-DECIGO and MAGIS are set to $[0.01, 100]~\text{Hz}$, and $[0.01, 3]~\text{Hz}$, respectively.

The GW frequency range of BBH inspiral is also a crucial factor that determines the frequency range of GW data.
The lower limit of this range is given by
\begin{equation}
    f_\text{low} = 0.035\left(\frac{\mathcal{M}}{20~M_\odot}\right)^{-\frac{5}{8}}\left(\frac{t_\text{obs}}{1~\text{yr}}\right)^{-\frac{3}{8}}~\text{Hz}~,
\end{equation}
where $\mathcal{M}$ is the detector-frame chirp mass. The observation duration of GW inspiral $t_\text{insp}$ is set to one year in our work. However, it should be noted that the appropriate value of $t_\text{obs}$ may vary depending on the mission duration and duty cycle of the GW detectors, which are currently uncertain. The upper GW frequency limit of BBH inspiral is given by the inner-most stable circular orbit (ISCO) frequency
\begin{equation}
    f_\text{ISCO}=\frac{1}{6^{3/2}\pi M } =220\left(\frac{M}{20~M_\odot}\right)^{-1}~\text{Hz}~,
\end{equation}
where $M$ is the detector-frame total mass. The frequency range of GW data is determined by the overlapping range between $[f_\text{low},f_\text{ISCO}]$ and a detector frequency band.

\section{GW Parameter estimation}\label{sec:PE}

\subsection{Basics of parameter estimation}
In this section, we provide a brief overview of the GW parameter estimation. For simplicity, we assume that the detector noise $\tilde{n}$ is stationary and Gaussian. From the stationarity, the noise covariance between the frequencies $f$ and $f'$ is given by
\begin{equation}
    \langle \tilde{n}^*(f) \tilde{n}(f')\rangle = \frac{1}{2}S_n(f) \delta(f-f')\, ,
\end{equation}
where $S_n(f)$ is the (one-sided) noise PSD. From the Gaussianity, the probability distribution of the noise $\tilde{n}(f_j)$ at the frequency bins $(f_0=f_i\, , f_1\, , \cdots ,\,f_{N-1}=f_e )$ follows (up to a normalization factor)
\begin{equation}\label{eq:noiseprob}
    p(\tilde{n}) \propto \exp\left[-\sum_{j=0}^{N-1} \frac{2 |\tilde{n}(f_j)|^2}{S_{n }(f_j)}\Delta f \right]\simeq \exp\left[-\frac{1}{2}(n|n)\right]
\end{equation}
where $\Delta f=f_{j+1}-f_j$ which is uniform, and
\begin{equation}
    (g|h) \equiv 4 \text{Re} \int_{f_i}^{f_e} df \frac{\tilde{g}^*(f) \tilde{h}(f)}{S_n(f)}\, 
\end{equation}
is the noise-weighted inner product of the strain data $g$ and $h$. The initial and end frequencies, $f_i$ and $f_e$, are determined by the detector frequency band and GW source, as described in \Sec{sec:detector}.

The GW detection significance of strain data $d$ can be measured by the likelihood ratio between the signal-plus-noise hypothesis($d = n + h$) and the noise-only hypothesis($d = n$)~\cite{LIGOScientific:2019hgc}. Here, the GW waveform model $h$ depends on the model parameters $\bm{\theta}$. By \Eq{eq:noiseprob}, the logarithm of the likelihood ratio is given by
\begin{equation}
    \ln \Lambda(d|\bm{\theta}) = (d|h(\bm{\theta}))-\frac{1}{2}(h(\bm{\theta})|h(\bm{\theta}))\, .
\end{equation}
The marginalization of $\Lambda(d|\bm{\theta})$ over the $\bm{\theta}$ parameter space can be used as the optimal detection statistics. The marginalization result is proportional to the maximum value of $\Lambda(d|\bm{\theta})$ when it is informative~\cite{Romano:2016dpx}. Since $ \Lambda(d|\bm{\theta})$ and $\ln \Lambda(d|\bm{\theta})$ are simultaneously maximized, the (maximum) matched-filter signal-to-noise ratio (SNR)
\begin{equation}
    \rho_m^2(d) \equiv 2 \max_{\bm{\theta}} \ln \Lambda(d|\bm{\theta}) 
\end{equation}
can be used as a measure of confidence in GW detection. 
Parameter $\bm{\theta}$ includes an overall amplitude $A$, the coalescence phase $\phi_c$, the coalescence time $t_c$, and the physical parameters $\bm{\mu}$ such as total mass and mass ratio. $\rho_m(d)$ can be analytically maximized with respect to $A$ and $\phi_c$ and becomes
\begin{equation}
    \rho_m^2(d) =  \max_{t_c,\bm{\mu}} \frac{1}{\rho^2_o(h)}\left[(d|h(t_c,\bm{\mu}))^2+(d|ih(t_c,\bm{\mu}))^2 \right]\, ,
\end{equation}
where $\rho_o^2(h) \equiv (h(\bm{\mu})|h(\bm{\mu}))$ is the optimal SNR.

The posterior probability distribution function (PDF) $p(\bm{\theta}|d)$ is obtained by the framework of Bayesian parameter estimation. Under a signal-plus-noise hypothesis, Bayes' theorem gives 
\begin{equation}\label{eq:posterior}
    p(\bm{\theta}|d) = \frac{p(\bm{\theta}) p(d|\bm{\theta})}{p(d)}\, ,
\end{equation}
where the logarithm of $p(d|\bm{\theta})$ is given by \Eq{eq:noiseprob} 
\begin{equation}\label{eq:lnp}
\begin{split}
    \ln p(d|\bm{\theta})&= -\sum_{j=0}^{N-1} \frac{2 |\tilde{d}_j-\tilde{h}(f_j;\bm{\theta})|^2}{S_{n }(f_j)}\Delta f +\text{const.}\\
    &\simeq -\frac{1}{2}(d-h(\bm{\theta})|d-h(\bm{\theta}))+\text{const.}
\end{split}    
\end{equation}
(up to an irrelevant constant). $p(\bm{\theta})$ is the prior PDF. For simplicity, we always use the uniform prior PDF in this work. 
The evidence $p(d)$ is given by
\begin{equation}\label{eq:evidence}
    p(d) = \int d\bm{\theta} p(d|\bm{\theta}) p(\bm{\theta})\, ,
\end{equation}
and it is a normalization factor for the posterior PDF.

\subsection{Computation of posterior PDF} \label{sec:numerical}

In contrast to the simple statement of Bayes' theorem, numerical computation of posterior PDF has many technical challenges when model parameter space has a high dimension. Several numerical algorithms have been developed to tackle the problems. We obtain posterior PDF using \textsc{Dynesty}~\cite{nested,nested2,speagle2020dynesty} sampler implemented in \textsc{Bilby}~\cite{bilby_paper} library. \textsc{Dynesty} sampler adopts the nested sampling algorithm~\cite{nested,nested2} which generates the posterior samples as a byproduct of computing the evidence \Eq{eq:evidence}. 

Parameter estimation in the decihertz band presents an additional challenge due to a large amount of GW phase evolution accumulated throughout a one-year-long observation. To accurately evaluate the likelihood \Eq{eq:lnp}, it is necessary to resolve the GW phase $\Psi(f) \propto \mathcal{M}^{-5/3}f^{-5/3}$ with sufficiently fine frequency bins. Note that the frequency resolution required to capture phase evolution $\Delta \Psi$ can be estimated by $\Delta f \propto \Delta \Psi \mathcal{M}^{5/3}f^{8/3}$. This estimation suggests that frequency binning in the decihertz band needs to be $10^8$ times finer than in the hectohertz band, resulting in a significant increase in the computing cost of the likelihood. 

To overcome the numerical challenges, we use the \textit{relative binning} method~\cite{relbin_paper, Zackay:2018qdy} for the likelihood computation. The relative binning method is based on the fact that the phase evolution difference of two slightly different waveforms changes slowly with frequency. Therefore, once the likelihood function is evaluated at a reference parameter, the likelihood function at nearby parameters can be evaluated by an interpolating function on coarser frequency bins. In this way, the relative binning achieves uniform computation time for the likelihood function regardless of frequency band and GW sources.

\subsection{Approximate covariance estimation}

One of the goals of constructing a posterior PDF is to obtain covariance between the parameters. Without using complicated numerical methods, sometimes an analytic covariance estimation can be useful as a crude estimation for an actual posterior PDF. Note that \Eq{eq:posterior} can be expanded around the maximum \textit{a posteriori} (MAP) $\hat{\bm{\theta}}$ as
\begin{equation}\label{eq:lnpexpand}
\begin{split}
    \ln p(\bm{\theta}|d) \simeq &\text{ const.}+ \frac{1}{2}\Delta \theta^a\Delta \theta^b \partial^2_{ab}\ln p (\hat{\bm{\theta}}|d)\\
    &+\mathcal{O}(\Delta\theta^3)
\end{split}    
\end{equation}
where $\Delta \bm{\theta} = \bm{\theta}-\hat{\bm{\theta}}$, and $\partial_a$ is the partial derivative with respect to $\theta^a$. We used $\partial_a\ln p=\partial_a p/ p=0$ at $\bm{\theta}=\hat{\bm{\theta}}$. The second derivative term is written as
\begin{equation}\label{eq:ddlnp}
\begin{split}
     \partial^2_{ab} \ln p(\hat{\bm{\theta}}|d) =&-\Gamma_{ab}(\hat{\bm{\theta}})\\
     & + (d- h(\hat{\bm{\theta}})|\partial^2_{ab}h(\hat{\bm{\theta}}))\, ,
\end{split}    
\end{equation}
where 
\begin{equation}
    \Gamma_{ab}(\bm{\theta})\equiv(\partial_a h(\bm{\theta})|\partial_b h(\bm{\theta}))
\end{equation}
is the Fisher information matrix (see Refs. \cite{Finn:1992wt,Finn:1992xs} for more rigorous treatment). When $d - h(\hat{\bm{\theta}})=n \ll h(\hat{\bm{\theta}})$ and $h(\bm{\theta})$ well approximates the true GW signals, the second term of the right-hand side of \Eq{eq:ddlnp} can be ignored. Note that $\mathcal{O}(\Delta \theta^3)$ terms in \Eq{eq:lnpexpand} is $(h|h)^{-1/2}\sim\text{SNR}^{-1}$ times smaller than $\mathcal{O}(\Delta \theta^2)$ terms. Therefore, at the high SNR limit, the posterior PDF can be approximated to the multivariate Gaussian distribution near $\hat{\bm{\theta}}$, and the covariance matrix $\Sigma$ can be obtained by
\begin{equation}\label{eq:covapprx}
    \Sigma^{ab}(\hat{\bm{\theta}})= (\Gamma^{-1})^{ab}(\hat{\bm{\theta}})\, .
\end{equation}
The $1\sigma$ statistical error of parameter $\theta^a$ is approximately $\sigma_{\theta^a} \simeq \sqrt{\Sigma^{aa}(\hat{\bm{\theta}})}$.

\section{Systematic error estimation}\label{sec:sys}

\begin{figure}
\includegraphics[width=\linewidth]{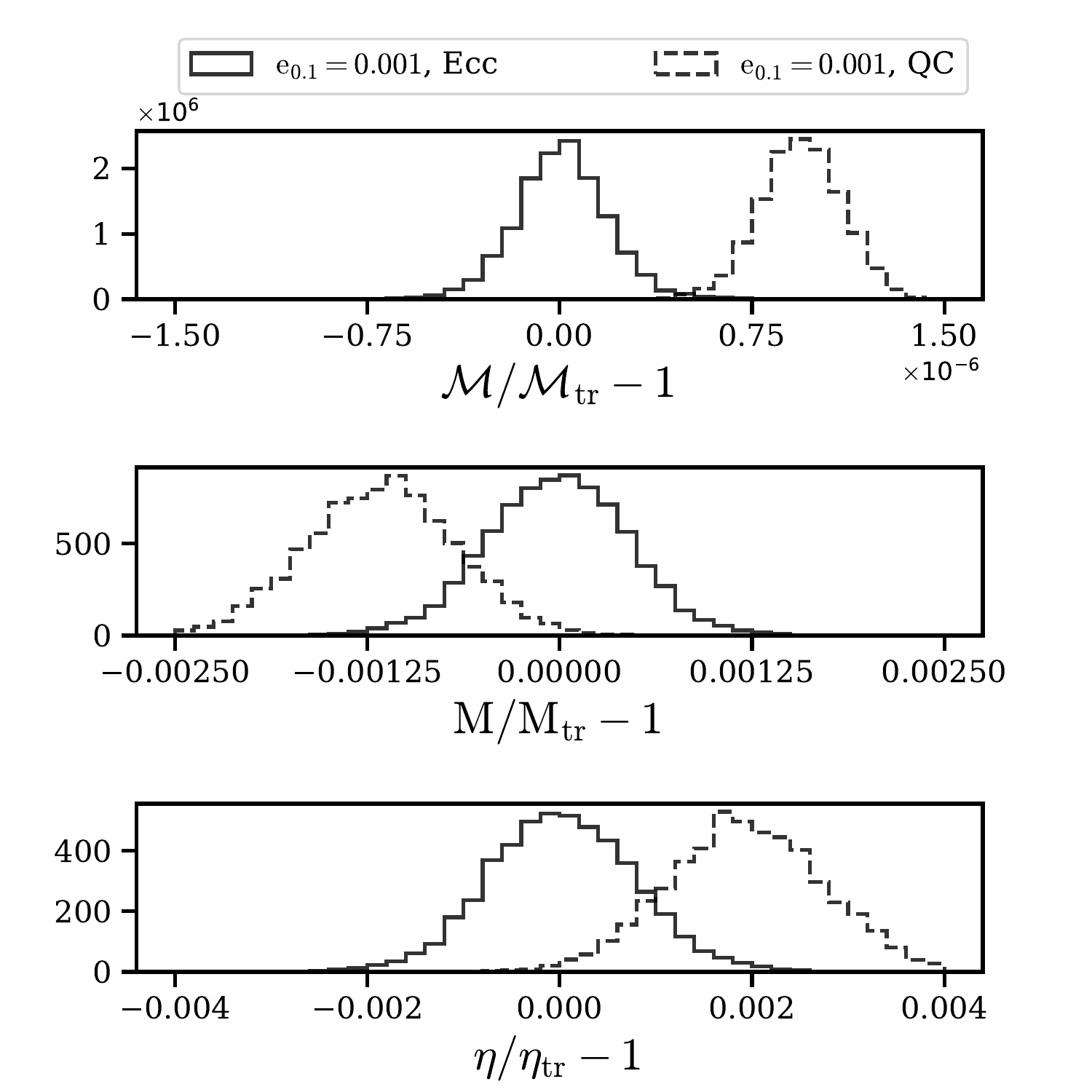}
\caption{ \label{fig:PDFex} 
 Posterior PDFs of eccentric GW signal computed by the correct model (TaylorF2Ecc, solid) and the inaccurate model (QC, dashed). The GW signal is generated by TaylorF2Ecc with $M_\text{tr}=40 \Msun$, $\eta_{tr}=0.24$, $e_{0.1}=0.001$, and $\overline{D}=5194\, \text{Mpc}$. We simulate the parameter estimation assuming one-year observation in B-DECIGO. This setup produces $\rho_o=30$. 
}
\end{figure}

\subsection{MAP shift and systematic error}
\label{sec:sysMAP}

When the waveform model $h(\bm{\theta})$ faithfully represents true GW signals $h_\text{tr}(\bm{\theta}_\text{tr})$, the posterior PDF is expected to peak around the true parameter of the signal $\bm{\theta}_\text{tr}$, which means that $\hat{\bm{\theta}}\simeq \bm{\theta}_\text{tr}$. However, the MAP $\hat{\bm{\theta}}$ does not always coincide with the true parameter due to the detector noise $n$ and waveform modeling error $\delta h = h_\text{tr}- h$. As long as the noise has a zero mean, the shift due to the noise will be averaged out, and only the bias due to the modeling error will remain. Therefore, denoting the shift of the MAP as $\Delta \theta^a$, a systematic error can be defined by
\begin{equation}\label{eq:sysdef}
    \Delta \theta^a_\text{sys} \equiv \langle \Delta \theta^a \rangle\, ,
\end{equation}
where $\langle \cdot \rangle$ means averaging over many noise realizations. In this work, we set $n=0$ and simply obtain $\Delta \theta^a_\text{sys} \simeq \Delta \theta^a$ by computing the posterior PDF only once, which is a good approximation for high SNR cases.

Figure \ref{fig:PDFex} shows the example of the MAP shift due to inaccurate waveform modeling. Here, we simulate a GW signal having $e_0=0.001$ with TaylorF2Ecc and compute the posterior PDF with the same waveform model (solid) and the QC model (dashed). The MAP shifts due to the use of the inaccurate(QC) model are clearly shown. While nontrivial distortion of the posterior PDF can arise (e.g., the biased posterior PDF having bimodality \cite{Cho:2022cdy}), we focus on the systematic error represented by the MAP shift.

\subsection{Fitting factor method}
\label{sec:sysFF}

MAP can be approximately obtained as a byproduct of finding the fitting factor(FF) instead of constructing a whole posterior PDF, which is computationally expensive. We will refer to this method as the FF method. FF is defined as the maximized match, where the match between $g$ and $h$ is the normalized inner product 
\begin{equation} \label{eq:match}
    \langle g | h \rangle \equiv \frac{( g | h ) }{(g|g)^{\frac{1}{2}} (h|h)^{\frac{1}{2}}}\, .
\end{equation}
Then FF can be written as
\begin{equation}\label{eq:FF}
    \text{FF} \equiv \max_{\bm{\theta}} \langle g|h(\bm{\theta}) \rangle\, .
\end{equation}
It is clear that $\text{FF}=1$ only if $h(\bm{\theta})$ is identical to $g$, otherwise $\text{FF}<1$. Mismatch(MM) defined as
\begin{equation}
    \text{MM} \equiv 1 - \text{FF}\, ,
\end{equation}
is also a useful measure of the difference between two waveforms.

Maximizing the match with respect to the constant phase can be done analytically, and \Eq{eq:FF} becomes
\begin{equation}\label{eq:FF2}
    \text{FF} = \max_{t, \bm{\mu}} \sqrt{\langle g|h (t,\bm{\mu})\rangle^2+\langle g |i h (t,\bm{\mu})\rangle^2 }\, .
\end{equation}
The right-hand side of \Eq{eq:FF2} is equal to $\rho_m(g)/\rho_o(g)$, implying that FF can be a measure of waveform template efficiency or, equivalently, SNR loss. Combined with the fact that $\rho\propto \overline{D}^{-1}$, inaccurate waveform modeling leads to GW event loss $1-\text{FF}^3\simeq 3 \text{MM}$ assuming the uniform-in-volume GW source distribution. Conventionally, an approximate waveform model is required to be at least $\text{FF}>0.97$ with respect to a more accurate waveform model, which corresponds to $\sim 90\%$ detection efficiency.

To estimate the systematic bias, we compute the match between $d=h_\text{tr}(\bm{\theta}_\text{tr})$ and $h(\bm{\theta})$, and maximize it with respect to $\bm{\theta}$. The maximum point of the match approximately coincides with the MAP. It can be shown using the fact that the derivative of the match can be written as
\begin{equation}
  \partial_a \langle d|h(\bm{\theta}) \rangle = \frac{(d-h|\partial_a h)-(d-h|h)(h|\partial_a \ln \mathcal{A} h)/(h|h)}{(d|d)^{\frac{1}{2}}(h|h)^{\frac{1}{2}}}\, .
\end{equation}
In the expression, $(d-h|\partial_a h)$ is dominant since it contains $\partial_a \Psi$, which is much larger than $\partial_a \ln \mathcal{A}$ in the other term. This is due to the large phase evolution of the GW chirps. Therefore, at the maximum point of the match, we can write
\begin{equation}\label{eq:FFpost}
\begin{split}
    \partial_a \langle d|h(\bm{\theta}) \rangle &\simeq \frac{(d-h|\partial_a h)}{(d|d)^{\frac{1}{2}}(h|h)^{\frac{1}{2}}} \\
    & = \frac{\partial_a \ln p(\bm{\theta}|d)}{(d|d)^{\frac{1}{2}}(h|h)^{\frac{1}{2}}} = 0\, ,
\end{split}
\end{equation}
which indicates that the maximum point of the match is also the MAP at the same time. 

Although computing $\text{FF}$ is less expensive than constructing the posterior PDF, the difficulties coming from long inspiral observation (as described in \Sec{sec:numerical}) still remain. We find that the relative binning method is very effective even for FF computation in the decihertz band. This is because the relative binning method basically realizes fast $(d|h(\bm{\theta}))$ computation.

\subsection{First order approximation and FCV method}
\label{sec:sysFCV}
The shift of MAP from $\bm{\theta}_\text{tr}$ can be estimated by using the local analytic properties of a posterior PDF~\cite{Flanagan:1997kp}. Assuming that MAP is at the stationary point of $p(\bm{\theta}|d)$, we have
\begin{equation}\label{eq:MAP}
    \partial_a \ln p(\hat{\bm{\theta}}|d) = (d - h(\hat{\bm{\theta}})|\partial_a h(\hat{\bm{\theta}})) =0 \, .
\end{equation}
Writing $\hat{\bm{\theta}}=\bm{\theta}_\text{tr} + \Delta \bm{\theta}$ and expanding \Eq{eq:MAP} up to the first order of $\Delta \bm{\theta}$ results in
\begin{equation}\label{eq:MAP1st}
(d - h(\bm{\theta}_\text{tr})|\partial_a h(\bm{\theta}_\text{tr})) - \Delta \theta^b \Gamma_{ab}(\bm{\theta}_\text{tr}) \simeq 0 \, .
\end{equation}
Here, we dropped $\Delta\theta^b (d-h(\bm{\theta}_\text{tr})|\partial_{ab}h(\bm{\theta}_\text{tr}))$ which is subdominant. Putting 
\begin{equation}
    d = n + h_\text{tr}(\bm{\theta}_\text{tr})=n + h(\bm{\theta}_\text{tr}) + \delta h (\bm{\theta}_\text{tr})\, 
\end{equation}
into \Eq{eq:MAP1st} and using $\Sigma^{ab}=(\Gamma^{-1})^{ab}$, we have
\begin{equation}\label{eq:bias}
\begin{split}
    \Delta\theta^a \simeq & \Sigma^{ab}(\bm{\theta}_\text{tr}) (n|\partial_b h(\bm{\theta}_\text{tr}))\\
    &+\Sigma^{ab}(\bm{\theta}_\text{tr}) (\delta h(\bm{\theta}_\text{tr})|\partial_b h(\bm{\theta}_\text{tr}))\, .
    \end{split}
\end{equation}
From the equation above, it is clear that the noise can shift MAP in a random direction, but the shift is zero on average. In contrast, the shift by the modeling error is not stochastic. Therefore, from \Eq{eq:sysdef}, we get 
\begin{equation} \label{eq:sysapprx}
    \Delta \theta^a_\text{sys} = \langle \Delta \theta^a \rangle\simeq  \Sigma^{ab}(\bm{\theta}_\text{tr}) (\delta h(\bm{\theta}_\text{tr})|\partial_b h(\bm{\theta}_\text{tr}))\, .
\end{equation}

Although \Eq{eq:sysapprx} provides a simple estimation for systematic error, it is not practical since what we obtain from the posterior PDF is $\hat{\bm{\theta}}$, not $\bm{\theta}_\text{tr}$. Instead, Ref. \cite{Cutler:2007mi} shows that systematic error can be estimated by replacing $\bm{\theta}_\text{tr}$ with $\hat{\bm{\theta}}$ in \Eq{eq:sysapprx}:
\begin{equation}\label{eq:FCV1}
     \Delta \theta^a_\text{sys} \simeq \Sigma^{ab}(\hat{\bm{\theta}}) (\delta h(\hat{\bm{\theta}})|\partial_b h(\hat{\bm{\theta}}))\, .
\end{equation}
When the modeling error is small, we can use
\begin{equation}\label{eq:FCV2}
    \delta h(\hat{\bm{\theta}}) \simeq \mathcal{A}_\text{tr} - \mathcal{A} + i \mathcal{A}(\Psi_\text{tr}-\Psi)\, .
\end{equation}
In this expression, the waveform amplitudes ($\mathcal{A}_\text{tr}$ and $\mathcal{A}$) and phases ($\Psi_\text{tr}$ and $\Psi$) are evaluated at $\hat{\bm{\theta}}$. This way of systematic error estimation is the Fisher-Culter-Vallissneri(FCV) method~\cite{Cutler:2007mi}.

Although the FCV method provides a simple one-step estimation for systematic errors, it should be utilized carefully when the assumptions behind this method do not hold. First of all, the FCV method can be inaccurate in large $\delta h$ situations, which is the limitation of the linear-order approximation. We find that the possibility of large $\delta h$ due to eccentricity is not negligible, and it will be discussed in \Sec{sec:sysresult}. Additionally, the method becomes inaccurate when MAP is located near the boundary points of the parameter ranges. This is because the FCV method relies on the stationary point condition \Eq{eq:MAP} which may not hold on the boundary points. However, we find that such cases are rare in the midband detectors due to the highly localized posterior PDF around MAP.

\section{Systematic error due to eccentricity}\label{sec:sysresult}
\subsection{Eccentricity scale of significant systematic error }

\begin{figure}
\includegraphics[width=\linewidth]{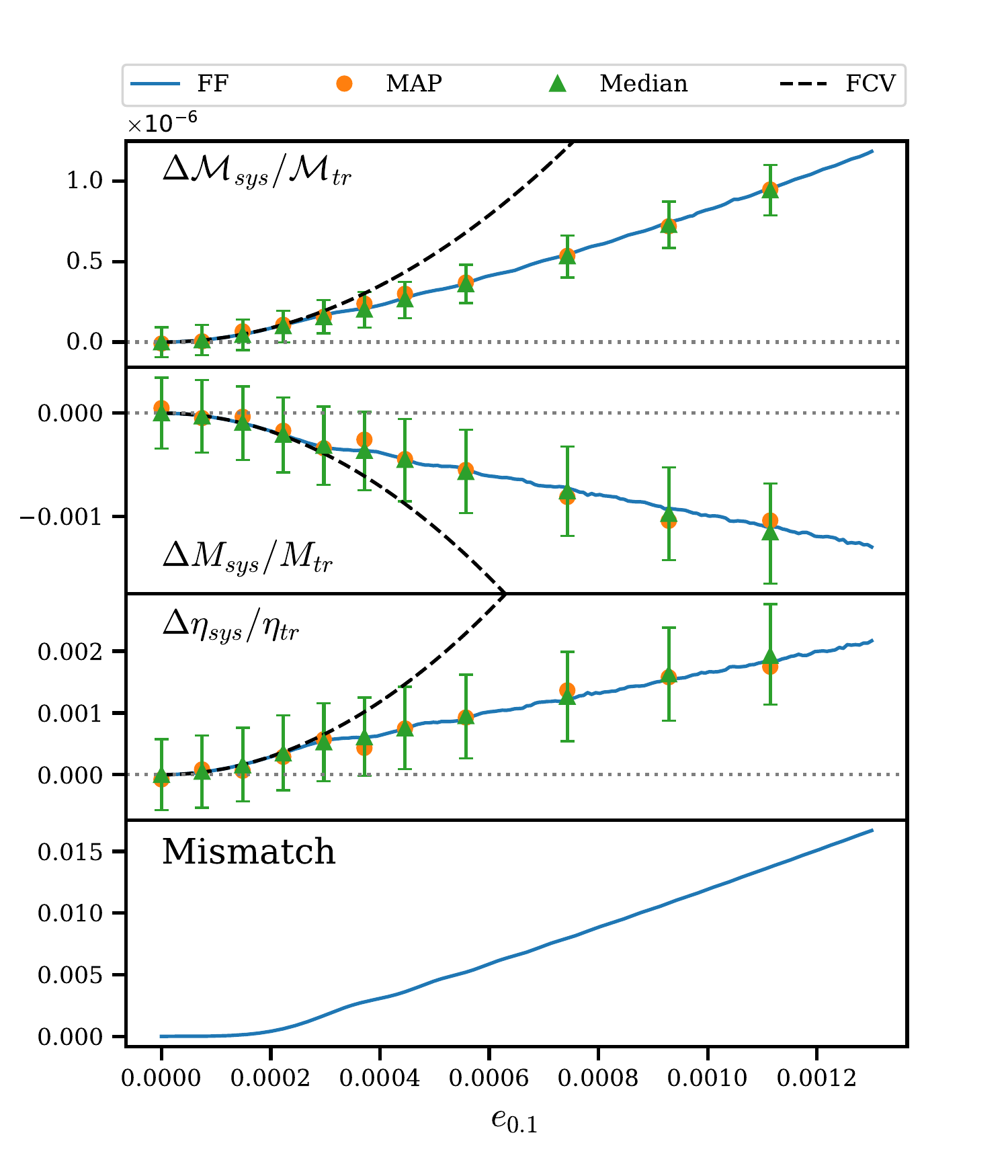}
\caption{ \label{fig:comparison1} 
First three panels from the top: systematic errors of $\mathcal{M}$, $M$, and $\eta$ as a function of $e_{0.1}$. The parameter estimation is done with the QC for eccentric GW signals. The GW signals are generated with $M=40\, \Msun$, $\eta=0.24$, and $\rho_o = 30$ while $e_{0.1}$ is varied in each simulation. The MAP shift (orange circles) and $1\sigma$ error bars (green bars) are obtained from the resulting posterior PDFs. The results are normalized by the true parameters of the GW signals. The error bars are drawn with respect to the (biased) median values (green triangle) which coincide well with the MAPs. Systematic error estimations with the FF method (blue solid curves) and the FCV method (black dashed curves) are also shown. Bottom panel: mismatch ($1-FF$) between the QC and the eccentric GW signals as a function of $e_{0.1}$.
}
\end{figure}

In the decihertz band GW observation, the quasicircular waveform model can induce non-negligible systematic errors even for GW signals with very small eccentricities. To demonstrate this, we generate eccentric GW signals and compute the posterior PDFs with QC assuming B-DECIGO observation. We obtain the MAP shifts and $1\sigma$ error ranges from the posterior PDFs. Figure \ref{fig:comparison1} summarizes the results for $\mathcal{M}$, $M$, and $\eta$ as a function of $e_{0.1}$. The figure shows that the MAP shifts (orange dots) of the parameters are already as large as $1\sigma$ error bars (green bars) at $e_{0.1}\sim 0.0004$. Since this level of eccentricity is not unlikely in the decihertz band\cite{Kowalska:2010qg,Rodriguez:2016kxx}, the results imply that the eccentric waveform model will be crucial for the precision test of GR in the decihertz band. In contrast to the large systematic errors, the mismatches(bottom panel of \Fig{fig:comparison1}) are only $0.01$ overall indicating a $1\%$ loss of SNR. This shows that the quasicircular waveform is still effective for GW detection purposes. The implications of these results in observational aspects will be discussed more in \Sec{sec:implication}.

Figure \ref{fig:comparison1} shows that the systematic error of chirp mass is more significant than those of total mass and mass ratio. For example, at $e_{0.1}\sim 0.0005$, the MAP shift of the chirp mass can be $3\sigma$ while that of the others are $1\sigma$. This is because the chirp mass controls the leading order(0 PN) behavior of GW phase evolution. Since total mass and mass ratio appear in the GW phase from the 1 PN corrections, they have smaller effects.

\begin{figure}
\includegraphics[width=\linewidth]{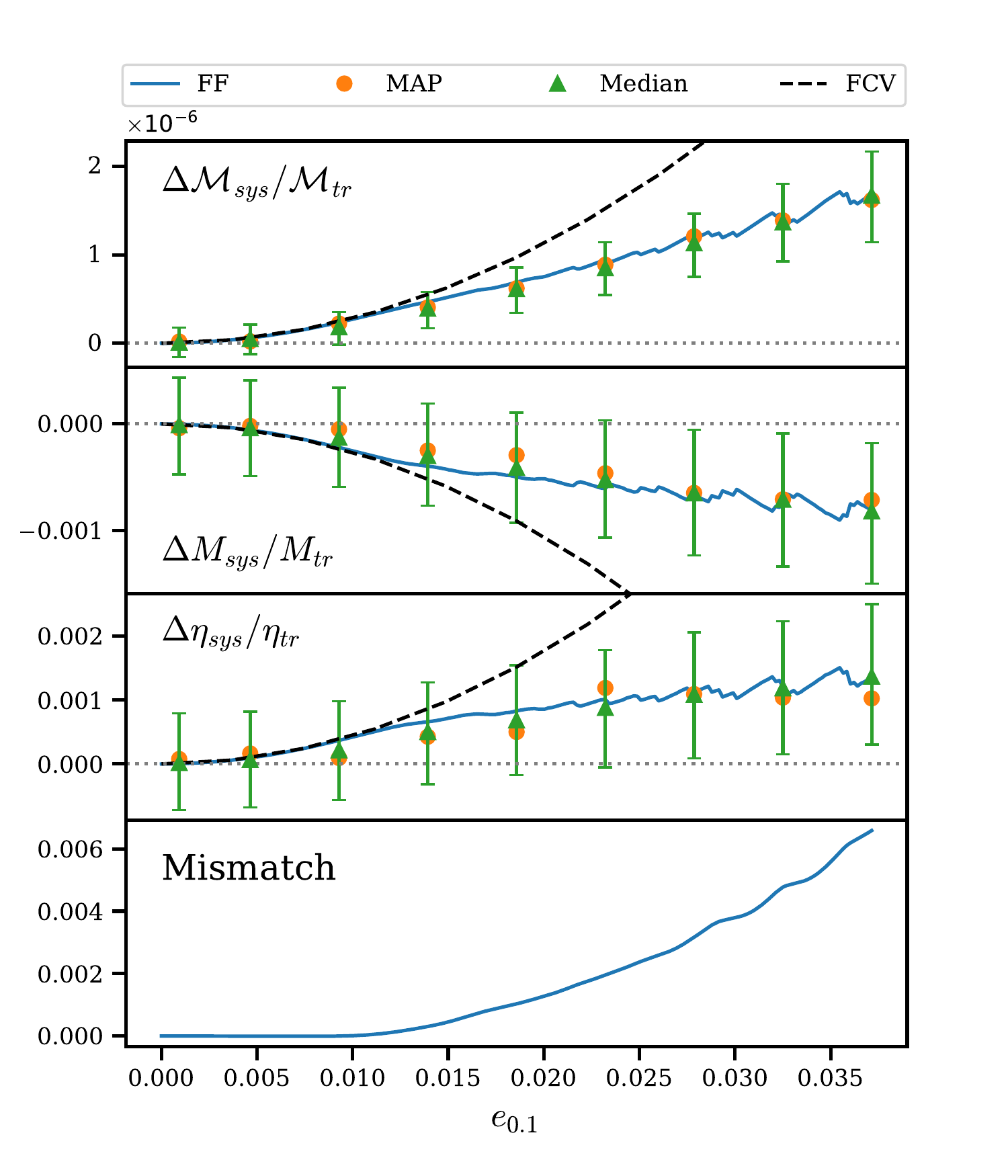}
\caption{ \label{fig:comparison2} 
Similar to Fig. \ref{fig:comparison1}, except Ecc0PN is used for parameter estimation. 
}
\end{figure}

We find that Ecc0PN has greatly enhanced accuracy compared to the QC. We compute the systematic errors of Ecc0PN by the same method as the QC case and summarize the results in \Fig{fig:comparison2}. It shows that the systematic errors of Ecc0PN become comparable with those of QC only when $e_{0.1}>0.01$. Also, the mismatch values of Ecc0PN are greatly reduced compared to those of QC. The effectiveness of Ecc0PN implies that the higher-order PN corrections have minor effects. Indeed, we find that the systematic errors of EccMPN ($M\geq 1$) are comparable to the QC and Ecc0PN cases only when $e_{0.1}\sim 0.1$ cases. This aspect will be discussed further in the next section.

\subsection{Validity of the approximate methods}

In Figs. \ref{fig:comparison1} and \ref{fig:comparison2}, we can see that the systematic error estimation with the FF method is consistent with the MAP shift of the actual posterior PDFs. This means that the approximation in \Eq{eq:FFpost} works well in our results. Therefore, the FF method is highly efficient in finding MAP. In \Sec{sec:implication}, We will use this method to compute systematic errors in much broader parameter space. Note that the MAP estimation with the FF method is effective only within the scope of our work, where the inaccuracy in the GW phase induces systematic errors. If the inaccuracy of GW amplitude modeling is non-negligible(for example, due to the higher harmonics), the FF method may not be appropriate for the MAP estimation. 

Estimations of the FCV method(dashed curves in Figs. \ref{fig:comparison1} and \ref{fig:comparison2}) are consistent with the other methods when $e_{0.1}$ is small, where the MAP shifts are within the $1\sigma$ error bars. However, as $e_{0.1}$ increases, deviations from other methods are observed, with this trend becoming more pronounced with higher $e_{0.1}$.  
The inaccurate estimations arise when $\Delta \Psi^\text{ecc}(f_e)-\Delta \Psi^\text{ecc}(f_i)> 6\pi$.
This means that the small modeling error assumption in \Eq{eq:FCV2} is not valid in this regime, and a higher-order calculation is required.

\section{Implications on detection and measurement}\label{sec:implication}

\begin{figure*}[ht]
\includegraphics[width=0.9\linewidth]{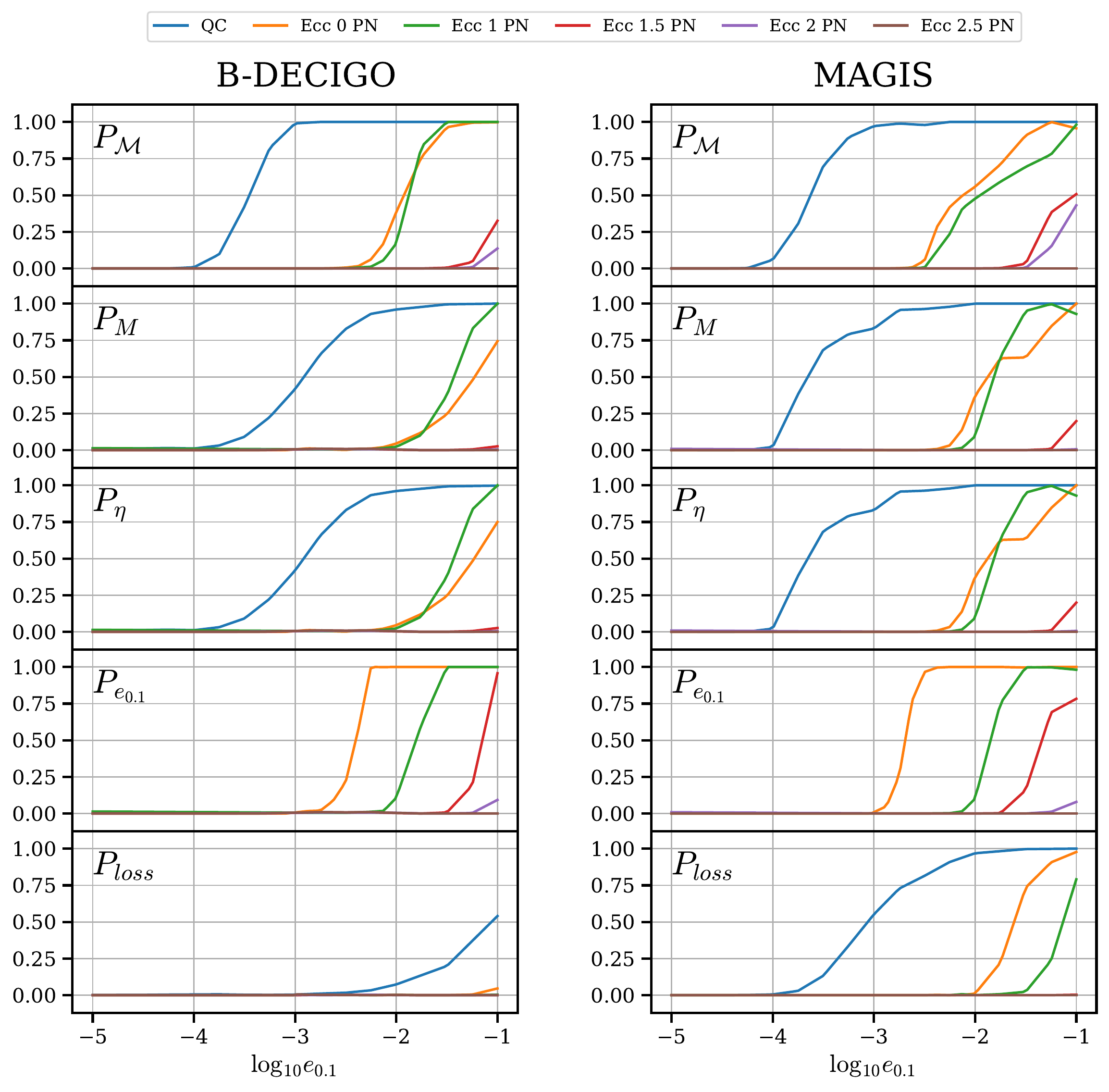}
\caption{ \label{fig:syslossdist} 
Biased measurement fractions $P_\mathcal{M}$, $P_M$, $P_\eta$, $P_{e_{0.1}}$ (the first four rows from the top), and event loss fraction $P_\text{loss}$ (the bottom row) caused by inaccurate waveform modeling in the eccentricity. The results are given a function of $e_{0.1}$, where those for different accuracy levels (QC and EccMPNs) are differentiated by the color of the curves. The left panels are the B-DECIGO cases, and the right panels are the MAGIS cases. Since QC has no $e_{0.1}$ parameter, there is no result for QC in the fourth row.
}
\end{figure*}

Inaccurate waveform modeling can cause systematic error and, in severe cases, lead to the loss of a GW merger event due to severe SNR loss. The degree of systematic error and SNR loss($\propto$ mismatch) have complicated dependence on GW parameters like black hole masses, luminosity distance, and eccentricity. To take into account the various factors, we simulate the BBH merger population using \textit{Power-law+Peak} (PP) model~\cite{Talbot:2018cva}. The model parameters of PP are set to the measurement values in Ref.~\cite{KAGRA:2021duu}. GW source redshifts are sampled uniformly in comoving volume up to $z=10$. We assume a constant source-frame merger rate. We collect the samples satisfying the detection criteria $\rho_o>8$ from the generated event samples. A total of 1024 samples are collected in each detector. 

Systematic errors $\Delta \theta_\text{sys}$ and the matched filter SNR $\rho_m $ are computed for the collected samples. The quantities are estimated by the FF method (\Sec{sec:sysFF}) for a given waveform model and an eccentricity value. We repeat the computations for eccentricities in the range $[10^{-5},10^{-1}]$.

We classify the event samples into several subsets. An event sample is regarded as detection if $\rho_m > 8$; otherwise, it is classified as loss. If an event sample is detection satisfying $\Delta \theta_\text{sys}^a >\sigma_{\theta^a}$, it is classified as biased. We use the Fisher information matrix to estimate $\sigma_{\theta^a}$. After the classification, we can define event loss fraction $P_\text{loss}$ and biased measurement fraction $P_{\theta^a}$ as
\begin{align}
    P_\text{loss} &= \frac{N_\text{loss}}{N_\text{total}}\\
    P_{\theta^a} &= \frac{N_{\theta^a}}{N_\text{detection}}\, ,
\end{align}
where $N_\text{loss}$, $N_{\theta^a}$, and $N_\text{detection}$ are the number of loss, biased $\theta^a$, and detection samples, respectively. Here, $N_\text{total}=1024$.

Our results show that the systematic error of QC can be significant for $e_{0.1}\sim \mathcal{O}(10^{-4})$. In both detectors, $P_\mathcal{M}>0.5$ for $e_{0.1}>10^{-3.5}$. The systematic errors of $M$ and $\eta$ are less significant than that of $\mathcal{M}$, but $P_{M}$ and $P_\eta$ are still non-negligible for $\mathcal{O}(10^{-4})$ eccentricities. For $e_{0.1}>10^{-3}$, more than half of $M$ and $\eta$ measurement can be biased. Since BBH mergers with $\mathcal{O}(10^{-4})$ $e_{0.1}$ may not be rare~\cite{Kowalska:2010qg,Rodriguez:2016kxx}, our results imply that a quasicircular waveform model is not appropriate for a precision test of GR. In this case, including the leading-order eccentric phase correction in the waveform model can be very effective. In \Fig{fig:syslossdist}, the Ecc0PN cases have negligibly small $P_{\theta^a}$ and $P_\text{loss}$ for the $\mathcal{O}(10^{-4})$ $e_{0.1}$. More accurate models (EccMPN with $M\geq1$) can be considered depending on the required precision and eccentricity distribution of BBHs.

In the QC case, $P_\text{loss}$ of MAGIS is significant for eccentricities of $\mathcal{O}(10^{-4})$, while that of B-DECIGO is negligible up to $e_{0.1}\sim 10^{-2}$. The difference can be explained by the difference in their sensitivity curve shape. In the case of B-DECIGO, the maximum sensitivity of B-DECIGO is reached at $\mathcal{O}(1)$ Hz(see \Fig{fig:sensitivity}). It means that $\mathcal{O}(1)$ Hz components of a GW signal drive parameter estimation results. In the case of MAGIS, however, decihertz components of a GW signal contribute to parameter estimation due to its (relatively) flat noise curve. Since an eccentric BBH has a larger eccentricity at a lower frequency, the phase difference between the QC and eccentric waveforms is also larger at a lower frequency. As a consequence, MAGIS becomes more sensitive to eccentricity and has higher $P_\text{loss}$. 

Higher-order EccMPN does not always improve accuracy. When Ecc0PN and Ecc1PN are compared, Ecc1PN leads to smaller $P_{e_{0.1}}$ and $P_\text{loss}$. However, the other bias probabilities can be worse(e.g. $P_M$ and $P_{\eta}$ results) at $e_{0.1}>0.01$. This implies that the PN approximation of eccentric waveform does not converge at 1 PN order for $e_{0.1}>0.01$. We find that the convergent behavior of accuracy appears only after 1.5 PN or higher orders. In other words, the high-order PN corrections are crucial for $e_{0.1}>0.01$.

From our $P_\text{loss}$ results, we can estimate the PN order requirement of the eccentric waveform model for GW detection. In B-DECIGO, Ecc1PN achieves nearly zero $P_\text{loss}$ for $e_{0.1}<0.1$. In MAGIS, the corresponding model is Ecc1.5PN. Depending on the eccentricity distribution of BBH, less accurate waveform models can be considered. For example, if the majority of BBHs are $e_{0.1}<10^{-2}$, then the QC waveform model can be adopted in B-DECIGO, although it takes the risk $P_\text{loss}\sim 0.1$. In the case of MAGIS, Ecc0PN can be considered when such eccentricity distribution is expected. 

The PN order requirement to avoid systematic error is more stringent. We find that, in both detectors, only Ecc2.5PN can maintain all of $P_{\theta^a}$s below $0.001$ for $e_{0.1}<0.1$. If BBH mergers with $e_{0.1}>10^{-2}$ are rare, Ecc1.5PN can be a practically good option for parameter estimation.

\section{Summary and Conclusion}\label{sec:conclusion}

We examined the influence of eccentricity on the detection and measurement in the decihertz detectors, B-DECIGO and MAGIS. If a waveform model is not accurate for eccentric binaries, systematic error can be a problem in the estimation of parameters from GW observations. The systematic error tends to increase as eccentricity increases. Since the expected eccentricity of BBHs is larger at lower frequencies, the systematic error problem can be significant in the midband detectors. Furthermore, the high precision of the midband detectors makes the problem more severe.

We studied the significance of the systematic error using the parameter set generated by the fiducial model of the BBH merger population. We found that even a small eccentricity, $e_{0.1}\sim\mathcal{O}(10^{-4})$, can cause systematic errors that are comparable to statistical errors if a quasicircular model is used. The risk of systematic error is greatly reduced by including the leading-order eccentric phase corrections. The higher-order corrections are important when $e_{0.1}>0.01$. To obtain negligible systematic error for $e_{0.1}>0.01$, the eccentric phase corrections up to 2.5 PN orders are required.

The PN accuracy requirement for GW detection is less stringent than for parameter estimation. In the case of B-DECIGO, the waveform model is still effective for detecting GW signals with $e_{0.1}<0.01$. If the eccentric phase corrections up to 1 PN order are included, B-DECIGO can get negligible event loss for $e_{0.1}<0.1$. In the case of MAGIS, the PN corrections up to 1.5 PN orders are required.

Our work can be the basis of more general parameter estimation studies in the decihertz detectors. In our work, we consider only the dominant harmonics of the GW waveform, but, in principle, the multiple harmonics induced by eccentricity can affect parameter estimation. Ultimately, studies on modeling accuracy for the case when black hole spin precession and eccentricity effects coexist are necessary. Although generating GW waveform for such studies is not trivial work, if it is possible, extending our work will be straightforward.


\begin{acknowledgments}
\bigskip

We thank Gihyuk Cho and Gungwon Gang for their helpful comments on our results. H.G.C. is supported by the Institute for Basic Science (IBS) under the project code, IBS-R018-D3. T.Y. is supported by “the Fundamental Research Funds for the Central Universities” under reference No. 2042024FG0009. H.M.L is supported by the National Research Foundation (NRF) of Korea Grants No. 2021R1A2C2012473 and No. NRF-2021M3F7A1082056.

\end{acknowledgments}

\appendix









\end{document}